\documentclass[journal=nalefd,manuscript=letter,layout=traditional]{achemso}

\usepackage[version=3]{mhchem} 
\usepackage[]{graphicx}
\usepackage{amsmath, amssymb}

\author{Sanghyun Jo}
\affiliation{DPMC and GAP, Universit\'{e} de Gen\`{e}ve, 24 quai Ernest Ansermet, CH-1211 Geneva, Switzerland}
\author{Nicolas Ubrig}
\affiliation{DPMC, Universit\'{e} de Gen\`{e}ve, 24 quai Ernest Ansermet, CH-1211 Geneva, Switzerland}
\author{Helmuth Berger}
\affiliation{Institut de Physique de la Matiere Complexe, Ecole Polytechnique Federale de Lausanne, CH-1015 Lausanne, Switzerland}
\author{Alexey B. Kuzmenko}
\affiliation{DPMC, Universit\'{e} de Gen\`{e}ve, 24 quai Ernest Ansermet, CH-1211 Geneva, Switzerland}
\author{Alberto F. Morpurgo}
\affiliation{DPMC and GAP, Universit\'{e} de Gen\`{e}ve, 24 quai Ernest Ansermet, CH-1211 Geneva, Switzerland}
\email{Alberto.Morpurgo@unige.ch}

\title [] {Mono- and Bilayer WS$_2$ Light-Emitting Transistors}


\begin{document}








\begin{abstract}
We have realized ambipolar ionic liquid gated field-effect transistors based on WS$_2$ mono- and bilayers, and investigated their opto-electronic response. A thorough characterization of the transport properties demonstrates the high quality of these devices for both electron and hole accumulation, which enables the quantitative determination of the band gap ($\Delta_{1L}=2.14$ eV for monolayers and $\Delta_{2L}=1.82$ eV for bilayers). It also enables the operation of the transistors in the ambipolar injection regime --with electrons and holes injected simultaneously at the two opposite contacts of the devices-- in which we observe light emission from the FET channel. A quantitative analysis of the spectral properties of the emitted light, together with a comparison with the band gap values obtained from transport, show the internal consistency of our results and allow a quantitative estimate of the excitonic binding energies to be made. Our results demonstrate the power of ionic liquid gating in combination with nanoelectronic systems, as well as the compatibility of this technique with optical measurements on semiconducting transition metal dichalcogenides. These findings further open the way to the investigation of the optical properties of these systems in a carrier density range much broader than that explored until now.
\end{abstract}


Layered transition metal dichalcogenides (TMDs) are attracting interest in nanoelectronics, owing to the possibility
to exfoliate high-quality thin two-dimensional (2D) crystalline flakes out of bulk single crystals, analogously to what is done for graphene \cite{novoselov2005two}. Investigations of the opto-electronic properties of these 2D crystals have started recently, focusing on different semiconducting TMDs, and revealing important differences with the bulk parents compounds \cite{mak2010atomically, splendiani2010emerging, kuc2011influence, zhao2013evolution, zeng2013optical, gutierrez2013extraordinary, sundaram2013electroluminescence, xiao2012coupled, mak2012control, zeng2012valley, cao2012valley, wang2012electronics, mak2013tightly, jones2013optical, ross2013electrical, pospischil2013solar, baugher2013electrically, ross2013electrically}. Most notably, the band-gap is found to increase due to quantum confinement, and a transition from indirect to direct gap is observed for individual monolayers \cite{mak2010atomically, splendiani2010emerging, kuc2011influence, zhao2013evolution, zeng2013optical, gutierrez2013extraordinary, sundaram2013electroluminescence, wang2012electronics}, for which unusual and strong effect of spin-orbit interaction have also been identified \cite{xiao2012coupled, mak2012control, zeng2012valley, cao2012valley, wang2012electronics, mak2013tightly, jones2013optical}. It has become clear that semiconducting TMDs are a very interesting and promising class of materials for both the study of new physical phenomena and, possibly, for practical applications.

Next to the possibility to control the electronic properties by selecting the specific semiconducting compound
(the chemical formula of TMDs is MX$_2$ with M a transition metal atom and X = S, Se, or Te; in the most common semiconducting compounds M = Mo or W)\cite{wang2012electronics} or the thickness of the material, several aspects of TMDs are particularly promising. For instance, the absence of covalent bonds between layers should enable the realization of high-quality hetero-structures, softening constraints due to imperfect lattice matching of different TMD compounds \cite{shi2012van, yu2012vertically, georgiou2013vertical, geim2013van}. Also, all semiconducting TMDs investigated to date have shown an excellent ambipolar behavior, allowing transport of both electrons and holes \cite{pospischil2013solar, baugher2013electrically, ross2013electrically, podzorov2004high, Hwang2012transistors, bao2013high, das2013wse2, zhang2012ambipolar, braga2012quantitative, ye2012superconducting, zhang2013formation}. For semiconductors this is a key aspect, which is needed for the realization of devices such as solar cells or light-emitting diodes, and for the implementation of complementary metal-oxide-semiconductor (CMOS) technology to integrate large numbers of devices in complex circuits.

For electronic systems with atomic thickness, it is difficult to implement controlled chemical doping without causing serious degradation. The investigation of the opto-electronic properties of very thin ambipolar TMD layers usually exploits field-effect transistor (FET) structures \cite{mak2013tightly, jones2013optical, ross2013electrical, pospischil2013solar, baugher2013electrically, ross2013electrically}. It has been shown experimentally that ambipolar transport can be induced in FETs with conventional solid-state gate dielectrics \cite{pospischil2013solar, baugher2013electrically, ross2013electrically, podzorov2004high, Hwang2012transistors, bao2013high, das2013wse2}. In that case, however, dielectric breakdown seriously limits the range of carrier density that can be accessed. This issue becomes even more serious for thinner TMD layers, for which the band gap is larger\cite{mak2010atomically, splendiani2010emerging, kuc2011influence, zhao2013evolution, zeng2013optical, gutierrez2013extraordinary, sundaram2013electroluminescence, wang2012electronics}, and so are the required gate voltages. A viable alternative to conventional dielectrics is the use of an ionic liquid to replace the gate dielectric.

Here we fabricate mono- and bilayer WS$_2$ ionic liquid gated FETs, and investigate their opto-electronic properties. Transport measurements show that high-quality ambipolar devices can be realized, with low contact resistance, high on-off ratio, virtually ideal subthreshold slopes, and room-temperature mobility values reaching up to approximately 50 cm$^2$/Vs. As for ambipolar WS$_2$ transistors based on thicker crystals\cite{braga2012quantitative}, this high quality allows us to extract quantitatively the band gap from the FETs transfer curves. According to expectations, the monolayer gap ($\Delta_{1L}= 2.14$ eV) is larger than the bilayer one  ($\Delta_{2L}= 1.82$ eV), and both are significantly larger than the bulk band gap\cite{braga2012quantitative} ($\Delta_{bulk} = 1.4$ eV). We further show that the devices can be biased to achieve simultaneous injection of electrons and holes at opposite contacts, and that this results --for both mono- and bilayers-- in light emission from the position of the FET channel where electrons and holes recombine. We compare the energy of the lines in the electro-luminescence and photo-luminescence spectra to the gap values measured from ambipolar transport, and estimate the exciton binding energies. Our results show how ionic liquid gating is a powerful technique to investigate transport and optical properties of TMD layers of atomic thickness, giving access to a range of carrier density much broader than that explored so far.

Ionic liquid gated FET structures are by now well established. They have been initially introduced in organic FETs, because the extremely large gate capacitance (originating from the nm-scale thickness of the double layer forming at the interface with material to be gated) enables a drastic reduction of the operating gate voltages \cite{panzer2005low, shimotani2006electrolyte, ono2008high, xia2009comparison, ono2010high}. Subsequently, their use with inorganic materials has led to the observation of a number of fascinating phenomena, such as gate-induced superconductivity\cite{ye2012superconducting, ye2010liquid, bollinger2011superconductor, ueno2011discovery} and magnetism\cite{yamada2011electrically,shimamura2012electrical}. In combination with TMDs, the possibility to induce ambipolar transport and to accumulate a very large density of carriers (in excess of 10$^{14}$ cm$^{-2}$) both in the conduction and in the valence band has been demonstrated \cite{zhang2012ambipolar, braga2012quantitative, ye2012superconducting, zhang2013formation}. The very high quality of ionic liquid gated TMD FETs allows, among others, to extract quantitatively the magnitude of the band-gap, directly from the difference of the threshold voltage for electron and  hole conduction \cite{braga2012quantitative}. Despite all these promising results, the use of ionic liquid gates with TMD monolayer systems has so far not been reported.

Fig. 1a-d show optical microscope images of different mono- and bilayer WS$_2$ flakes, before and after attaching electrical contacts.
The flakes were produced by adhesive tape exfoliation of bulk crystals, which were subsequently transferred onto a Si/SiO$_2$ substrate (the crystals were grown by vapor phase transport; for details see reference \citenum{braga2012quantitative}). The thickness of the SiO$_2$ layer was chosen to be 90 nm, to optimize the optical contrast\cite{castellanos2010optical, benameur2011visibility} of thin flakes under the optical microscope. Through a systematic analysis of RGB images of many of these thin flakes, in conjunction with atomic force microscopy measurements, we calibrated the optical contrast in the R, G, and B channels, to enable the univocal identification of the flake thickness (see Fig. 1e). Flakes used in the device fabrication were selected using this technique and photoluminescene measurements (see Fig. 5) confirmed the validity of our identification. Contacts were patterned using conventional electron-beam lithography and lift-off and consisted of a 50 nm Au layer, annealed\cite{radisavljevic2011single} at 200 $^\circ$C for two hours in a  flow of Ar (100 sccm) and H$_2$ (10 sccm). In the same evaporation step, we also deposited a large Au pad acting as a gate (see Fig. 1f for a schematics of the device configuration). A small droplet of ionic liquid (IL) DEME-TFSI (\emph{N,N-diethyl-N-methyl-N-(2-methoxyethyl) ammonium bis (trifluoromethylsulfonyl) imide})\cite{zhang2012ambipolar,ye2012superconducting,zhang2013formation, ye2010liquid, bollinger2011superconductor, yamada2011electrically}, was placed onto the device in a glove-box with sub-ppm O$_2$ and H$_2$O concentration, with a top glass coverslip sealing and flattening the droplet, as it is needed to obtain a sufficiently sharp focal-point under a microscope. A silver wire treated with piranha solution was also inserted in the IL, and used as reference electrode, enabling us to measure the potential drop across the IL/WS$_2$, and to eliminate the effects of potential drops at the gate/IL interface. All transport and optical measurements were performed in cryostats or vacuum systems at a pressure below 10$^{-6}$ mbar (after transferring the device in the vacuum chamber of the measurement system we always waited for one day, to remove humidity and oxygen present in the IL).

We start with the electrical characterization of IL-gated  FETs. Figures 2a and 2b show the transfer curves (i.e., the source-drain current $I_{SD}$ as a function of the gate voltage  V$_G$) of mono- and bilayer devices, for different values of the applied source-drain voltage V$_{SD}$. Measurements were performed at temperatures not far above the freezing point\cite{zhang2013formation} of the IL, at 240 K for the monolayer device and at 230 K for the bilayer one. Clear ambipolar behavior with negligible hysteresis is seen in both devices. Despite the sizable band gap of WS$_2$ monolayer ($\sim$ 2.1 eV) and bilayer ($\sim$ 1.8 eV; WS$_2$ has the largest band-gap in the group of semiconducting-TMDs) \cite{kuc2011influence, zhao2013evolution,zeng2013optical, gutierrez2013extraordinary, wang2012electronics}, high on-state currents can be achieved both when accumulating electrons and holes, thanks to the high efficiency of the IL gating technique. Although most measurements were performed in a two terminal configuration, the devices had additional contacts enabling multi-terminal measurements which allow us to estimate the contact resistance for both electron and hole injection. For the largest charge accumulation shown in Fig. 2 the contact resistances are 4 k$\Omega$ (6.5 k$\Omega$) and 52 k$\Omega$ (12 k$\Omega$) for the electron- and hole-doped regions, in the monolayer (bilayer) device (the smaller current on the hole side observed in Fig. 2a, therefore, is likely due to a larger contact resistance. The difference in contact resistance values that we observe is likely a consequence of small, uncontrolled details in the device fabrication procedure. Indeed, through the analysis of many different devices, we have observed that the values of contact resistance can vary between few k$\Omega$ and few tens of k$\Omega$, irrespective of the mono-or bilayer nature of the device). In the off-state of the devices (i.e., when the gate is biased to position the Fermi level in the band gap), the measured current is much smaller than in the on-state (between 1 and 10 pA --see inset of Fig.2a,b-- corresponding to the background current of the measurement system), indicating a very low density of unintentional dopants in our WS$_2$ crystals. The resulting on/off ratio is between 10$^5$ and 10$^6$, with the precise value depending on the device and on whether electron or hole accumulation is considered (this value is limited by our sensitivity in measuring the off current).

The high quality of the mono- and bilayer FET characteristics enables us to extract the value of the band-gap, as we have previously done for bulk WS$_2$ \cite{braga2012quantitative}. For such a quantitative analysis, it is important to look at the electrical characteristic as a function of the voltage measured at the reference electrode ($V_{REF}$, which corresponds to the voltage effectively responsible for the accumulation of charge at the surface of WS$_2$) \cite{braga2012quantitative}. Fig. 3a and 3b show --in logarithmic scale-- the source drain current $I_{SD}$ as a function of $V_{REF}$ for mono- and bilayer, in the turn-on region of the device for both hole and electron conduction. The steep increase of the current with $V_{REF}$ correspond to value of subthreshold swing $S$ for electron and hole transport of  52 (63) and 57 (67) mV/dec for the  monolayer (bilayer) device. These data  nearly perfectly match the ultimate limit for the subthreshold swing $S = kT/e \ln(10)$ at the measurements temperature ($T=260$ K) \cite{sze2006physics}, which demonstrate the virtually complete absence of charge traps in the channel and at the contacts of our devices, as well as the ideal gate-coupling between the IL and the WS$_2$ FET channel.

Just like in the case of bulk WS$_2$\cite{braga2012quantitative}, it is the ideality of these IL gate devices that enables the quantitative determination of the gap directly from the difference of threshold voltages for electron and for hole conduction (the gap extracted from these  measurements corresponds to the smallest distance --in energy-- between the valence and conduction band, i.e. the smallest between the direct and the indirect gap in the system). To extract  the values of the threshold voltage for electrons and holes, $V_{TH}^e$ and $V_{TH}^h$, we extrapolate linearly the $I_{SD}$-vs-$V_{REF}$ curve to $I_{SD}=0$ in the electron and hole turn-on regions of the FETs, as shown in in Fig. 3c,d. We then obtain $\Delta V_{GAP}= V_{TH}^e- V_{TH}^h = 2.14$ V for the monolayer and 1.82 V for the bilayer. As we will discuss below, after having presented the results of the optical studies, these estimates of the gap ($\Delta_{1L}=2.1$ eV and $\Delta_{2L}=1.8$ eV)  appear to be fully consistent with the results of the photo- and electro-luminescence measurements.

To complete the electrical characterization of our devices, we estimate the value of the mobility of electrons and holes from the measured device conductivity $\sigma$ (since $\sigma$ is obtained from the conductance measured in a two-terminal configuration, which also contains the effect of the contact resistance, the resulting mobility values may be  underestimated). Starting from $\sigma = n_{e/h}e \mu_{e/h}$, with $n_{e/h}e=C_*(V_{REF}-V_{TH}^{e/h}) $ ($C_*$ is the capacitance per unit area of the ionic liquid) and $\mu_{e/h}$ the mobility of electrons and holes, we have that $\mu_{e/h}=\frac{1}{C_*} \frac{d \sigma}{d V_{REF}}$ (with this quantity calculated above the threshold voltage for electrons or holes, respectively).  Using the estimated values for the capacitance per unit area of the DEME-TFSI ionic liquid --7.2 and 4.7 $\mu$Fcm$^{-2}$ for electron and hole accumulation\cite{zhang2013formation}-- we obtain mobility values of 44 (19) and 43 (12) cm$^2$V$^{-1}$s$^{-1}$ for electron and hole carriers in monolayer (bilayer). These values are comparable with those reported in the literature for different few-layers and monolayer TMD semiconductors \cite{podzorov2004high, bao2013high, zhang2012ambipolar, braga2012quantitative, ye2012superconducting, zhang2013formation, radisavljevic2011single, jariwala2013band, radisavljevic2013mobility, baugher2013intrinsic}.

To realize light emitting transistors, we need to operate devices in the regime of simultaneous electron and hole injection at the the two opposite contacts (Fig. 1g). This requires applying a sufficiently large source-drain bias, to locally invert the potential of the channel with respect to the gate. The regime of simultaneous electron/hole injection has been investigated very extensively in organic light-emitting transistors\cite{meijer2003solution, rost2004ambipolar, dinelli2006high, zaumseil2006spatial, bisri2009high}: it is easy to identify experimentally, because it manifests itself in a very steep increase in $I_{SD}$ occurring at large $V_{SD}$, starting at the end of the conventional FET saturation regime. For the monolayer, this is shown in Fig. 4a,b, for the case in which the gate voltage at low $V_{SD}$ bias would result in the accumulation of electrons (Fig. 4a) or of holes (Fig. 4b; the bilayer exhibits qualitatively identical characteristics). When biasing the device in the region where the current increase occurs, light should be coming out of the device, which --as we show below-- is the case.

Before looking at the electroluminescence in the ambipolar injection regime, we analyze the photoluminescence signal, to check whether the ionic liquid --which completely covers the TMD layers-- does affect the light emitted by the device (it could absorb strongly, or generate a signal with spectral features in the frequency range where the TMD layers emit light). The photoluminescence signal due to the ionic liquid on top of the substrate, measured by shining a laser beam at an energy of $E = 2.54$ eV (488 nm) in a region near the device, is shown in Fig. 5a. It consists of a rather weak, smooth background, with a  sharp peak  at approximately 2.17 eV, which corresponds well to the energy expected due to the Raman shift (2973 cm$^{-1}$) associated with the stretch-mode of C$-$H groups in DEME-TFSI\cite{yoshimura2011pressure}. We take advantage of this sharp, identifying feature, to remove the background from the photoluminescence signal measured on the TMD devices. This we do by simply subtracting a signal proportional to the background, with proportionality constant fixed by the condition that the peak is not visible any more after subtraction. The background-subtracted, photoluminescence signals are shown in Fig. 5b for the mono- and bilayer, as well as for a thicker (bulk) layer of WS$_2$.

The position and intensity of the spectral features in the photoluminescence of the different layers agrees with literature results\cite{zhao2013evolution, zeng2013optical, gutierrez2013extraordinary}, which confirms our identification of mono- and bilayers. The monolayer exhibits a single, very intense peak, at $E \simeq 2.0$ eV, corresponding to transition associated to the direct band-gap. The bilayer and the bulk also exhibit a peak at a comparable energy, albeit of largely reduced intensity. They additionally exhibit a peak at lower energy (and lower for the bulk as compared to the bilayer), originating from optical transitions associated to the indirect gap. The energy of the different observed transitions is summarized in Fig. 5c, together with the value of the gap determined from the transport experiments (see discussion below). We conclude that the presence of the IL does not prevent the measurement of photoluminescence and should not therefore affect the observation of electroluminescence.

Fig. 6a,b  show optical microscope images of the monolayer device under two different bias conditions ($V_G = 0.1$ V and $V_{SD}$ = 2.6 V in a; $V_G = 0.1$ V and $V_{SD}$ = 2.9 V in b), both in the ambipolar injection regime. The Au electrodes used to contact the flake, as well as the thicker WS$_2$ flake on the side are clearly visible. Although the presence of the IL makes it harder to obtain sharp images on this small dimension scales, light emission --the red spot clearly apparent in both images-- can be seen directly under the optical microscope in both cases. The position of the emission spot is noticeably different in the two cases, as expected: indeed in the ambipolar injection regime the location in the FET channel where electrons and holes recombine can be shifted simply by changing the gate and source drain bias, as it has been beautifully demonstrated for organic light-emitting transistors\cite{zaumseil2006spatial, bisri2009high}. At the nano-scale, with a material of atomic thickness, it will be particularly interesting to investigate the mechanism determining the spatial extension of the emission region.

Fig. 6c shows the spectrum of the electro-luminescence signal measured for both the mono- and the bilayer device. As expected,
the electro- and the photo-luminescence spectra resemble each other. A direct comparison of the two spectra (for both mono- and bilayer), normalized to unity at the highest point, is shown in Fig. 6d,e, where it can be seen that the main peaks coincide in position. For the bilayer the signal originating from the indirect transition is relatively stronger in intensity, which is probably due to the fact that under the bias conditions needed to generate electro-luminescence, the device effective temperature (i.e., the effective temperature of charge carriers and phonons) is significantly higher than that of the environment. We also performed similar measurements for thicker (bulk) devices, but in that case no electro-luminescence was detected. This clearly shows that going to atomic thickness with TMD semiconductors allow new opto-electronic functionalities --possibly relevant for practical device applications-- not present in the bulk parent compounds.

Finally, we discuss the energy position of the spectral peaks in the observed optical transitions, summarized in Fig. 5c. For the monolayer, optical measurements only give access to transitions associated to the direct gap, whose energy is smaller than the gap value extracted by transport measurements. This is expected because of excitonic effects \cite{cheiwchanchamnangij2012quasiparticle, komsa2012effects, ramasubramaniam2012large}. For the bilayer, the direct-gap optical transition occurs at an energy much larger than the transport gap, which is also expected, since transport measures the smallest gap. Additionally, the transport gap is somewhat larger than the energy of the optical transition associated to the indirect gap, again a manifestation of excitonic effects\cite{cheiwchanchamnangij2012quasiparticle, komsa2012effects, ramasubramaniam2012large}. For the bulk, the transport gap and the energy of the indirect optical transition coincide within the precision of the measurements. We conclude that the hierarchy of the energies extracted from the optical transitions and from transport gap agrees with expectations. Quantitatively, the measurements allow us to estimate the magnitude of excitonic effects in WS$_2$ systems of different thickness by subtracting the optical transition energy from the value of the gap obtained from transport. We find a binding energy of 160 meV for the (direct) exciton in monolayers and of 80 meV for the bilayer (indirect) exciton. While these values are certainly affected by the dielectric environment through screening\cite{cheiwchanchamnangij2012quasiparticle, komsa2012effects, ramasubramaniam2012large} (and so results obtained on IL gated devices may differ quantitatively from devices with other types of dielectrics), our measurements reveal that the exciton binding energy decreases with increasing layer thickness. This trend can be explained qualitatively by considering that in a thicker flake the electron and hole can be further apart, resulting in a reduction of their mutual Coulomb interaction. A detailed theoretical analysis would be desirable to enable a quantitative comparison with the experimental data, from which information about the microscopic structure of the exciton may be obtained.

In summary, we have investigated the opto-electronic properties of atomically thin ionic-liquid gated light-emitting transistors realized on WS$_2$ mono- and bilayers. Our results show the power of the ionic liquid gating technique, which enables the determination of the band-gap through simple transport measurements, is compatible with optical measurements, and enables ambipolar injection in a simple device configuration, based on a single gate electrode. We further show how our data enable a quantitative estimate of the excitonic binding energies. These results open the way to the experimental investigation of the opto-electronic properties of atomically thin semiconducting transition metal dichalcogenides, in a much broader range of carrier density than what has been possible to access until now.

\begin{acknowledgement}
We are grateful to A. Ferreira, I. Gutierrez Lezama, J. Teyssier, and D. van der Marel for technical help, discussions, and support.
We gratefully acknowledge the Swiss National Science Foundation for financial support through
individual projects of AFM and ABK, as well as through a SNF Sinergia proposal on atomically thin transition metal dichalcogenides.
AFM and ABK further acknowledge financial support from the EU "Graphene Flagship" project.

\end{acknowledgement}





\bibliography{WS2_LE-FET_cond-mat}

\begin{figure}
  \includegraphics[width=.8\textwidth]{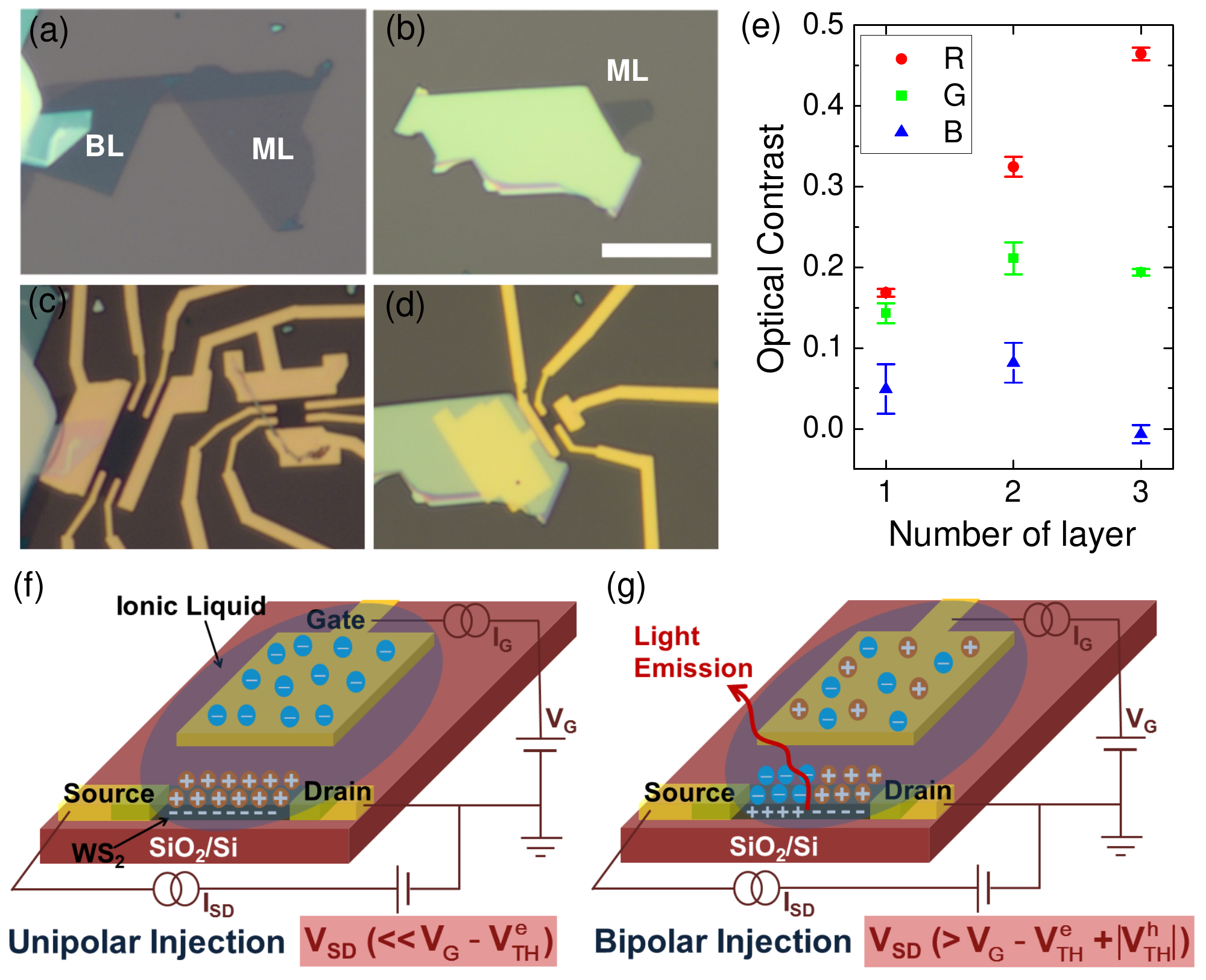}
  \caption{Optical microscope images of mono- (ML) and bilayer (BL) flakes are shown in (a) and (b), and the corresponding devices after attaching Au electrodes in (c) and (d) (the scale bar is 10 $\mu$m long, and is the same for all panels). Panel (e) shows the optical contrast in the R, G, B channels for mono-, bi-, and trilayer flakes. Each point corresponds to the average of the contrast measured on approximately five flakes (the error bars are determined by the standard deviation of these measurements). Panels (f) and (g) are schematic illustrations of an ionic liquid gated device, in the unipolar (f) and ambipolar (g) injection regime. The actual devices also contain a reference electrode not shown in these schematics.}
  \label{fgr1}
\end{figure}

\begin{figure}
  \includegraphics[width=.6\textwidth]{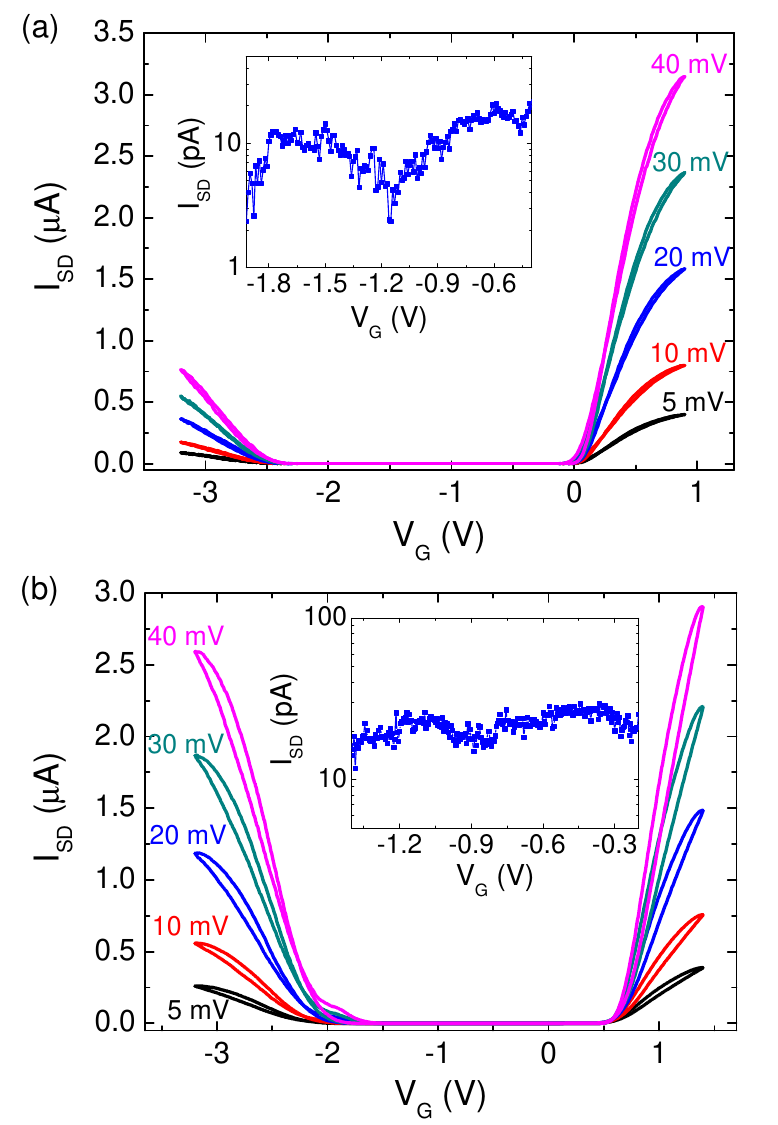}
  \caption{Transfer curves of a monolayer (a) and of a bilayer (b) device, exhibiting ambipolar behavior. The data show the $V_G$ dependence of $I_{SD}$ for various different values of source-drain voltage $V_{SD}$, as indicated in the figure. The smaller current on the hole side for the monolayer device in (a) is due to a larger contact resistance. Note that the hysteresis observed upon sweeping $V_G$ forward and backward is very small (as compared to most ionic liquid gated devices reported in the literature). The insets in the two panels zoom on the off current of the corresponding device, measured at $V_{SD}$ = 20 mV.}
  \label{fgr2}
\end{figure}

\begin{figure}
  \includegraphics[width=.8\textwidth]{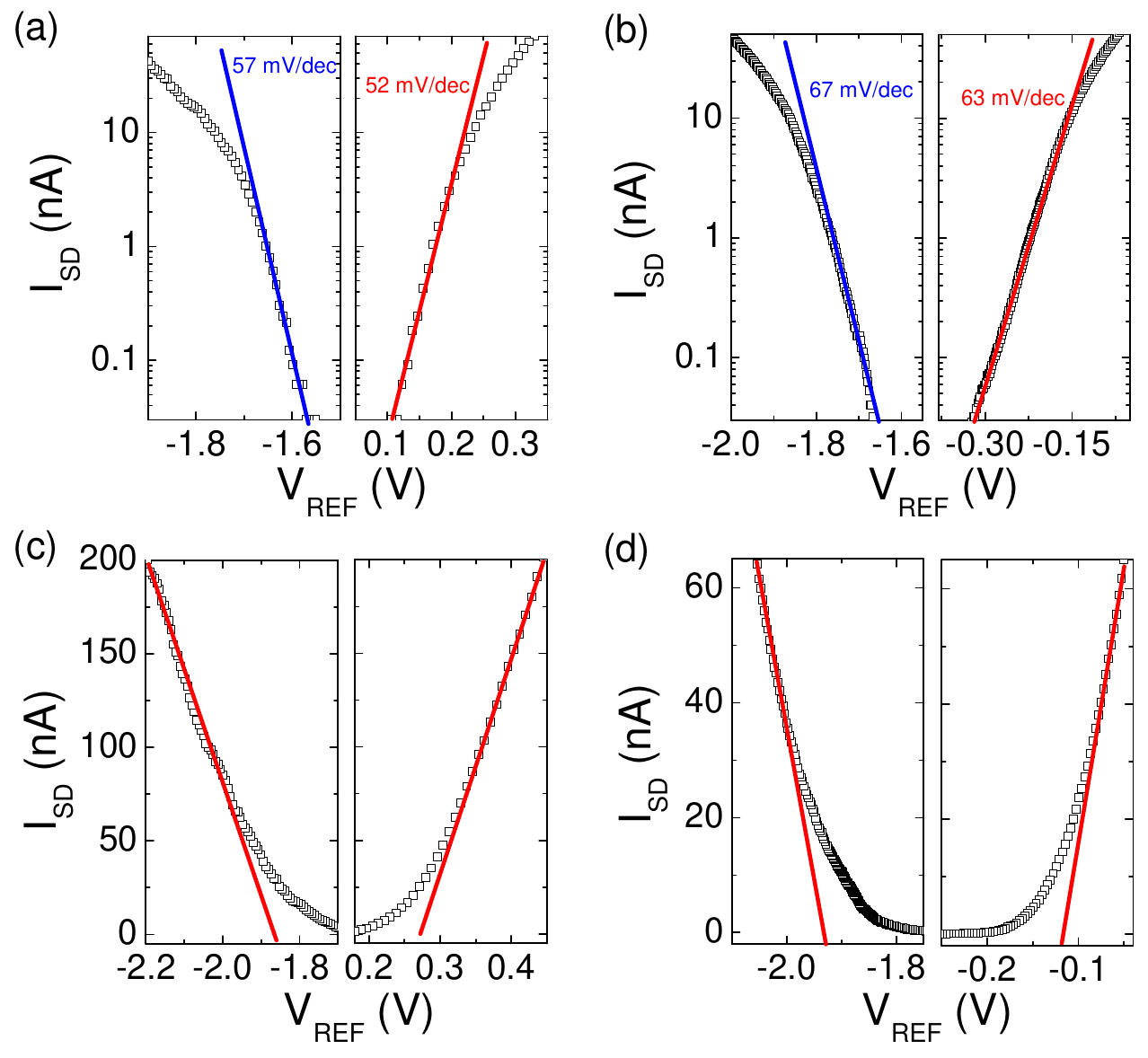}
  \caption{Log-scale plots of the source-drain current $I_{SD}$ as a function of reference voltage $V_{REF}$ for the monolayer (a) and bilayer (b) device, in the electron (left) and hole (right) turn-on region, measured at V$_{SD}$ = 5 mV. The solid lines are linear fits to the data,  from which we extract the values of the  subthreshold swings indicated in the figures. Panels (c) and (d) --for mono- and bilayer, respectively-- show the same data in linear scale. The solid lines linear extrapolations done to extract the values V$_{TH}^e$ and V$_{TH}^h$ of the electron and hole threshold voltages.}
  \label{fgr3}
\end{figure}

\begin{figure}
  \includegraphics[width=.6\textwidth]{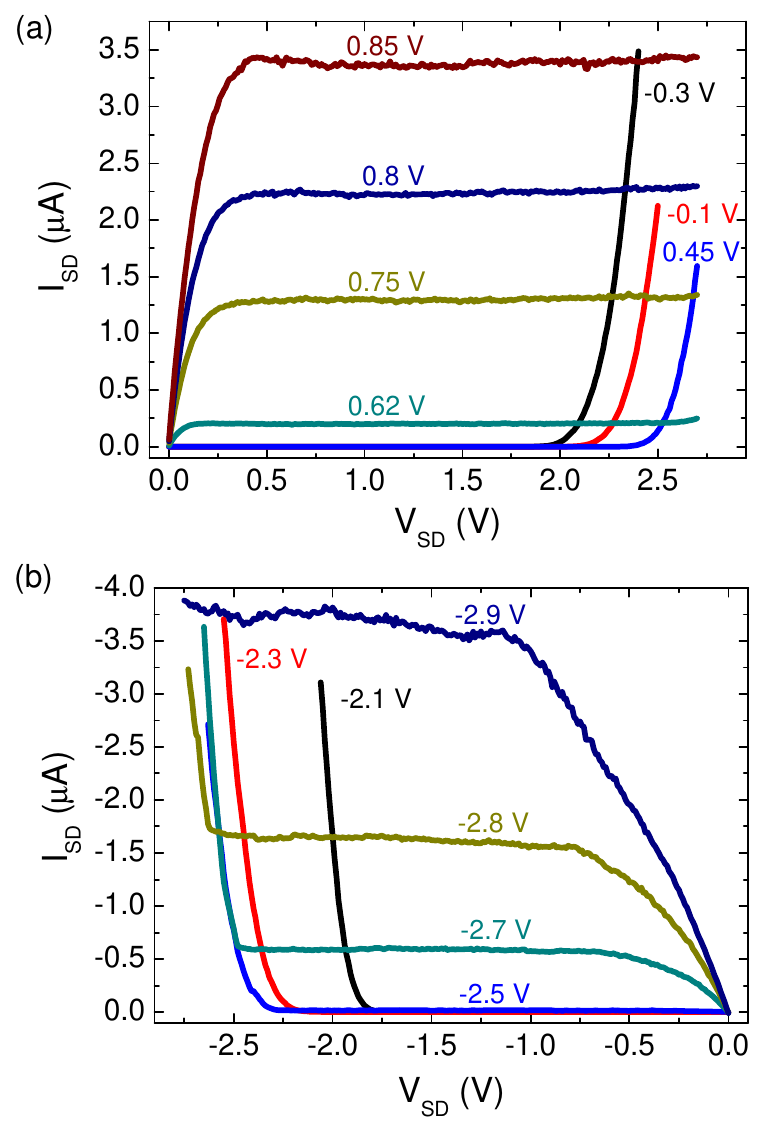}
  \caption{Evolution of the source-drain current $I_{SD}$ as a function of source-drain voltage $V_{SD}$ for different values of gate voltage $V_G$, in the monolayer device. At small $V_{SD}$, the gate voltage is set to form an electron channel in panel (a) and a hole channel in panel (b). Note the clear saturation regime starting from which the current steeply increases at sufficiently large values of $V_{SD}$, when the ambipolar injection regime is reached. It is in this ambipolar injection regime that light-emission is expected. The bilayer device shows virtually identical electrical characteristics.}
  \label{fgr4}
\end{figure}

\begin{figure}
  \includegraphics[width=.7\textwidth]{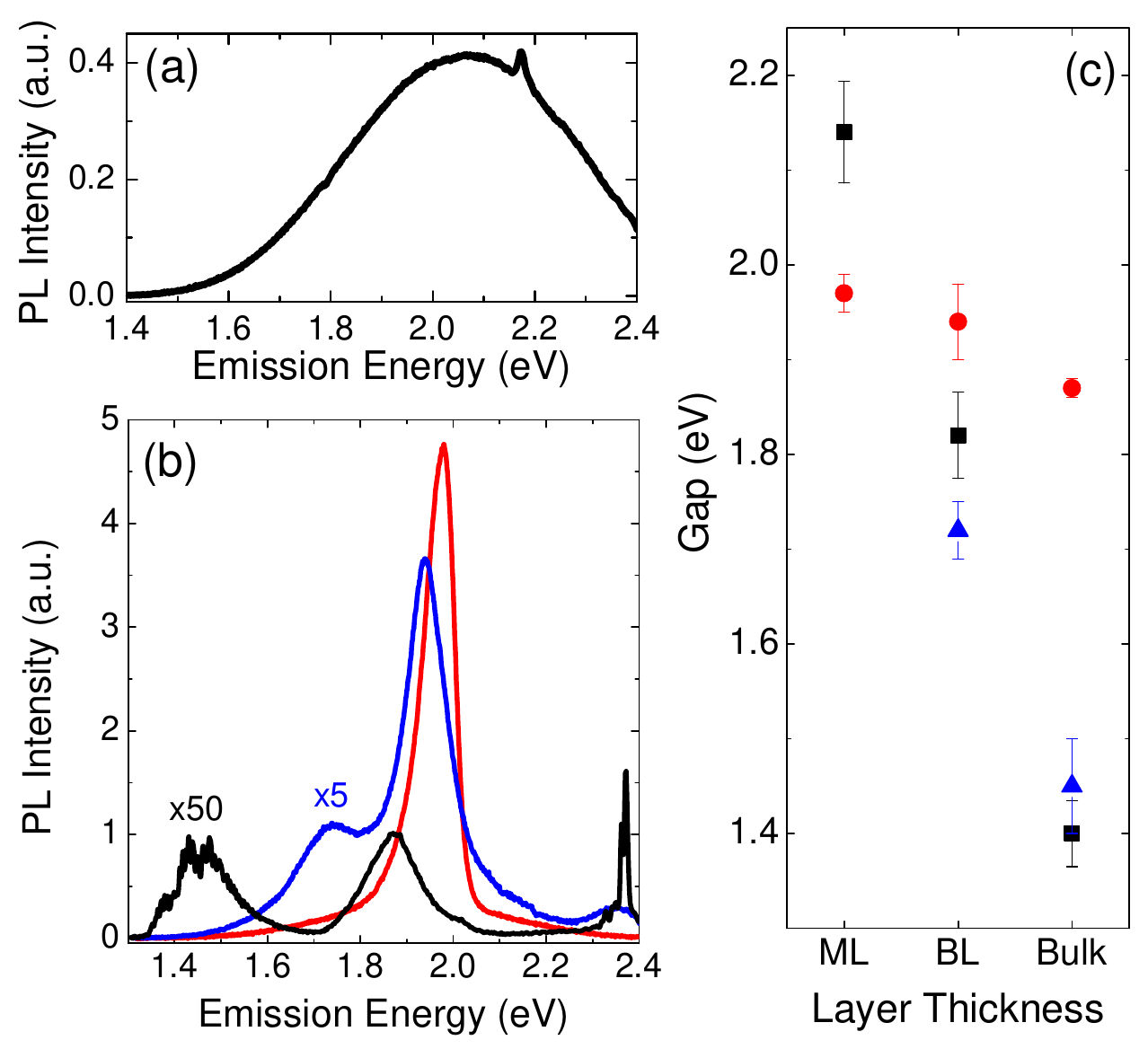}
  \caption{(a) Photoluminescence signal measured on the substrate covered with ionic-liquid, away from the WS$_2$ flake and from the gold contacts. The intensity of the background is much smaller than the photoluminescence signal measured on the mono- and bilayer flakes. It consists of a broad and featureless signal, except for the small sharp peak at 2.17 eV, due to Raman scattering from the C-H groups in the ionic liquid. (b) Background-subtracted photoluminescence signal for the monolayer (red line), the bilayer (blue line), and a thick (bulk) flake (black line). (c) Summary of the relevant energies measured by transport and photoluminescence on mono- and bilayers and on a bulk flake. The black squares correspond to the band gap values extracted from transport measurements. The red circles corresponds to the energy of the optical transition associated to the direct gap. The blue triangles represent the energy of the optical transition associated to the indirect gap.}
  \label{fgr5}
\end{figure}

\begin{figure}
  \includegraphics[width=.7\textwidth]{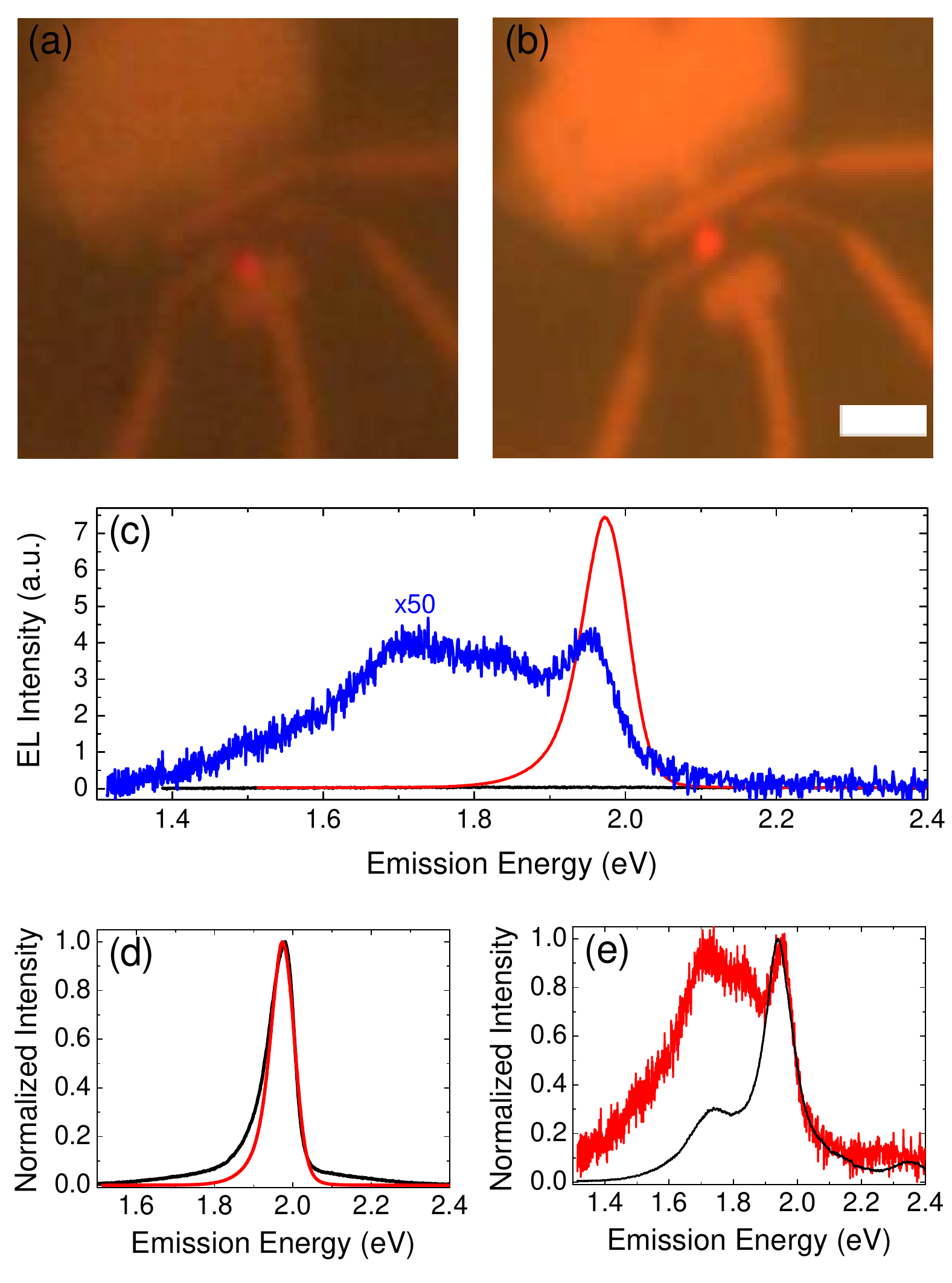}
  \caption{Optical microscope images of the monolayer device biased at V$_{SD}$ = 2.6 (a) and 2.9 V (b) (V$_G$ = 0.1 V in both cases), with clearly visible a spot of emitted light due to electro-luminescence. Note that the displaced position for the two different bias values, which is expected since the location of the recombination zone in a light-emitting transistor depends on the bias. The scale bar in (b) corresponds to 2 $\mu$m and is the same in both figures. (c) Electroluminescence spectrum for the monolayer (red line), bilayer (blue line) and bulk (black line). Panels (d) and (e) show a comparison of the normalized electro- and photo-luminescence spectra for the monolayer (d) and the bilayer (e). In both panels the red line corresponds to the electroluminescence signal and the black line to the photoluminescence one.}
  \label{fgr6}
\end{figure}

\end{document}